\documentclass[twocolumn,showpacs,superscriptaddress,preprintnumbers,amsmath,amssymb, prl]{revtex4}

\usepackage{graphicx}
\usepackage{dcolumn}
\usepackage{bm}
\usepackage{amssymb}

\newcommand{\coolingandso}{Effective radial and axial confinement frequencies are (32, 113) $\mathrm{Hz}$ [(28, 112) $\mathrm{Hz}$] for $^{87}$Rb and (44, 194) $\mathrm{Hz}$ [(31, 191) $\mathrm{Hz}$] for $^{40}$K, for lattice depth of $2$ ($17$)~$E_r$.}
\newcommand{\criticalratio}{Assuming a critical ratio of $U/J = 29.3$ ($49.8$) for bosonic filling 1 (2), see \cite{CAP07}.}

\begin{document}

\preprint{APS/123-QED}

\title{Role of interactions in $^{87}$Rb-$^{40}$K Bose-Fermi mixtures in a 3d optical lattice}

\author{Th.~Best}
\author{S.~Will}
\author{U.~Schneider}
\author{L.~Hackerm\"uller}
\affiliation{Institut f\"ur Physik, Johannes Gutenberg-Universit\"at Mainz, Staudinger Weg 7, 55128 Mainz, Germany}
\author{D.-S.~L\"uhmann}
\affiliation{I. Institut f\"ur Theoretische Physik, Universit\"at Hamburg, Jungiusstr. 9, 20355 Hamburg, Germany}
\author{D.~van Oosten}
\altaffiliation{Present address: FOM Institute for Atomic and Molecular Physics (AMOLF), Kruislaan 407,
1098 SJ Amsterdam, The Netherlands}
\author{I.~Bloch}
\affiliation{Institut f\"ur Physik, Johannes Gutenberg-Universit\"at Mainz, Staudinger Weg 7, 55128 Mainz, Germany}

\date{\today}

\begin{abstract}
We investigate the effect of interspecies interaction on a degenerate mixture of bosonic $^{87}$Rb and fermionic $^{40}$K atoms in a three-dimensional optical lattice potential. Using a Feshbach resonance, the $^{87}$Rb-$^{40}$K interaction is tuned over a wide range. Through an analysis of the $^{87}$Rb momentum distribution, we find a pronounced asymmetry between strong repulsion and strong attraction. In the latter case, the Bose-Hubbard parameters are  renormalized due to self-trapping, leading to a marked shift in the superfluid to Mott insulator transition with increasing Bose-Fermi interaction.
\end{abstract}


\maketitle

Mixtures of quantum gases in optical lattices form novel quantum many-body systems, whose properties are governed by the interplay of quantum statistics, inter- and intraspecies interactions, as well as the relative atom numbers of the components. In particular, degenerate Bose-Fermi mixtures, which rarely occur in nature and have only recently come within experimental reach, have stimulated theoretical investigations. 
Already for the simplest description of such an interacting mixture in terms of a single-band Bose-Fermi Hubbard model, a variety of quantum phases have been predicted \cite{ALB03, LEW04}, including charge-density-waves, polaron-like quasi-particles \cite{MAT04}, as well as perturbed Mott-insulating (MI) states \cite{CRA04, VAR07}. The correlated (or anti-correlated) movement of composite particles can result in exotic superfluids \cite{ZUJ07}.
Moreover, within a limited parameter regime, supersolid behavior has been predicted \cite{BUE03, TIT07}.
Unlike in the pure bosonic case, realistic descriptions of the mixture need to consider multi-band physics \cite{LUE07}, finite-temperature \cite{CRA08}, physical loss channels, and the inhomogeneity present in optical lattice experiments. Recently, mixtures of bosonic $^{87}$Rb and fermionic $^{40}$K atoms  have been investigated at fixed attractive background scattering length in two independent experiments \cite{GUE06, OSP06B}. The authors found a shift of the visibility loss of the $^{87}$Rb interference pattern towards shallower lattices, which has been interpreted as a shift of the bosonic superfluid to MI transition. Several mechanisms have been put forward to explain this shift, however, none could clearly be identified.
\begin{figure}[b!]
\includegraphics[width=8.4cm]{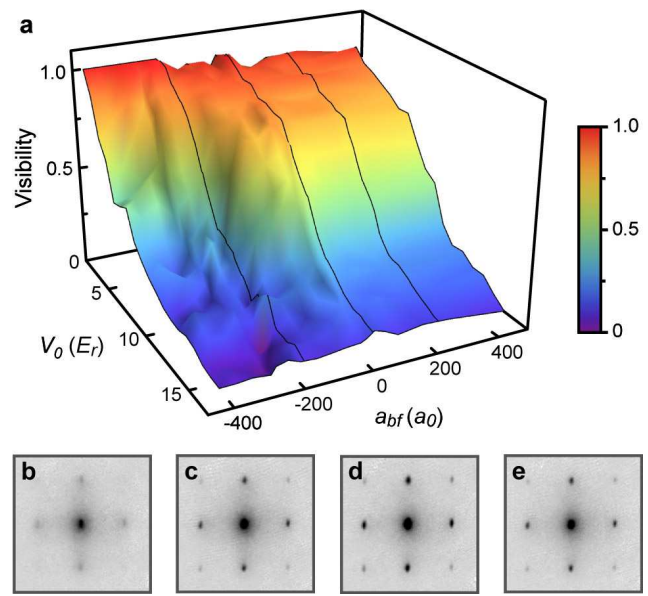}
\caption{\label{FIG1} Dependence of bosonic interference pattern on interspecies interaction. (a) Visibility of the $^{87}$Rb interference pattern versus lattice depth $V_0$ and interaction strength, parametrized by the interspecies scattering length $a_{bf}$, for high $N_f$ (interpolation from $1004$ individual measurements). (b)-(e) Exemplary TOF images of $^{87}$Rb atoms released from a $9\,E_r$ deep  optical lattice. The values for the interspecies scattering length are (b) $a_{bf} = -160\,a_0$, (c) $a_{bf} = -80\,a_0$, (d) $a_{bf} = +5\,a_0$, and (e) $a_{bf} = +235\,a_0$.}
\end{figure}

Here, we report on the realization of a quantum degenerate mixture of  bosonic $^{87}$Rb and fermionic $^{40}$K atoms with tunable interspecies interaction in a 3d optical lattice. We investigate the coherence properties of the bosonic component in the presence of fermions over a wide range of interspecies interactions, from strongly attractive to strongly repulsive, for varying admixture of fermions. At large fermion fillings and attractive interspecies interactions, we find good agreement with a recent model of self-trapping \cite{LUE07}. This leads to a significant renormalization of the Bose-Hubbard parameters, resulting in a shift of the superfluid to MI, in good agreement with our experimental results.

Starting point for our experiments is a degenerate mixture of $4\times 10^5$ $^{87}$Rb and up to $3\times 10^5$ $^{40}$K atoms in their respective hyperfine ground states $\vert F = 1, m_F = +1 \rangle$ and $\vert F=\frac{9}{2}, m_F = -\frac{9}{2}\rangle$. The mixture is held in a horizontally crossed pancake-shaped dipole trap, operated at a wavelength of $1030\,\mathrm{nm}$. The ellipticity of the dipole beams makes the trap vertically tight enough to ensure interspecies overlap in the presence of gravitational sag. 

A homogeneous magnetic field is applied in order to address the $2.9\,\mathrm{G}$ wide interspecies Feshbach resonance (FBR) at $546.9\,\mathrm{G}$  \cite{KLE07}. We can continuously tune the interspecies scattering length $a_{bf}$ between $-170\,a_0$ and $+800\,a_0$ below the FBR, and between $-800\,a_0$ and $-200\,a_0$ above the FBR, with an accuracy of $\pm 10\,a_0$ close to the resonance, where $a_0$ is the Bohr radius.
After $50\,\mathrm{ms}$ settling time for the magnetic field, a 3d optical lattice is adiabatically ramped up within $100\,\mathrm{ms}$, reaching a final depth in the range of $V_0=2$ to $17\,E_r$. The lattice wavelength $\lambda = 755\,\mathrm{nm}$ is blue-detuned with respect to the D2 transition of both $^{87}$Rb and $^{40}$K and chosen such that the lattice depths, as measured in units of the respective recoil energies $E_r = h^2 / (2 \, m_{b,f}\, \lambda^2)$, are equal for both species. This ensures maximum overlap of the Wannier functions, while the ratio of the tunneling rate of bosons and fermions is given by $m_f/m_b$. The blue lattice also causes an underlying harmonic anti-confinement which is (over-)compensated by the dipole trap \footnote{\coolingandso}. 

After a hold time of $100\,\mathrm{ms}$, which is long compared to the tunneling time for any lattice depth used, all traps as well as the magnetic field are instantaneously switched off, and the atom clouds are allowed to expand during $18\,\mathrm{ms}$ time of flight (TOF).
The resulting interference pattern of the $^{87}$Rb atoms is recorded using standard absorption imaging [see  Fig.~\ref{FIG1}(b)-(e)]. From these images, we extract the contrast in terms of the visibility \cite{GER05}. Furthermore, we determine the condensate fraction from a fit to the interference pattern.

\begin{figure}[b!]
\includegraphics{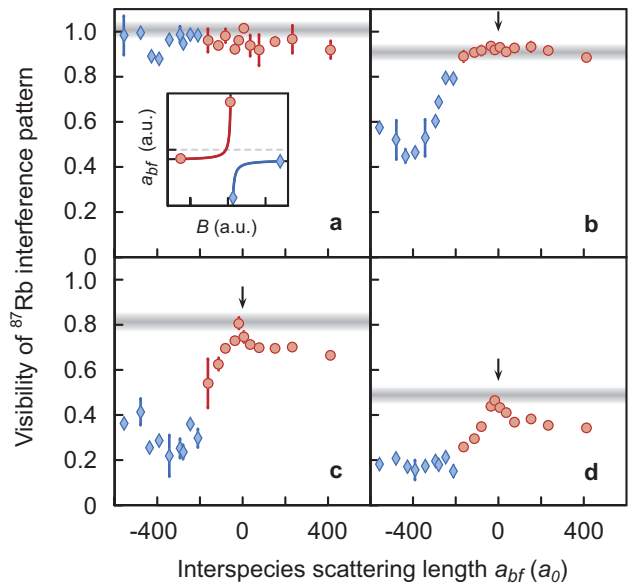}
\caption{\label{FIG2} Profiles of visibility versus interspecies scattering length at lattice depths of (a) $3\,E_r$, (b) $5\,E_r$, (c) $9\,E_r$ and (d) $12\,E_r$  for high $N_f$. At all lattice depths beyond $3\,E_r$, the visibility shows a maximum at a position consistent with $a_{bf}=0$ as indicated by the arrow within the calibration uncertainty. Blue diamonds (red circles) indicate points measured above (below) resonance, as depicted in the inset of (a). The shaded region represents the visibility measured for a pure $^{87}$Rb cloud using the same experimental parameters.}
\end{figure}

By varying both the lattice depth and the magnetic field, we probe the phases of the mixture. We have analyzed scenarios with low ($N_f\approx 0.25 N_b$), intermediate ($N_f \approx 0.5 N_b$) and high ($N_f \approx 0.75 N_b$) $^{40}$K atom numbers. The visibility diagram of the mixture with high $N_f$ is displayed in Fig.~\ref{FIG1}. Profiles of the same data are shown in Fig.~\ref{FIG2}, illustrating the most prominent features.

For shallow lattices of less than $3\,E_r$, we find a high visibility, which is almost independent of the interaction strength. Also for deeper lattices, at $a_{bf}\approx 0$, we recover a monotonic decay of the visibility versus lattice depth compatible with the superfluid to MI transition in a pure $^{87}$Rb sample \cite{GER05}. This demonstrates that the fermion cloud becomes fully transparent for the bosons. When the absolute value of the scattering length is increased, the visibility decay is shifted towards shallower lattices.  For small absolute values of $a_{bf}$, the decrease of visibility  is symmetrical around $a_{bf}=0$. The symmetry persists within the interval $\vert U_{bf}\vert \lesssim U_{bb}$, where the interboson repulsion is strong enough to effectively hinder bosonic double occupation of sites. Departing from this symmetry peak, we see a further decay of visibility, which is significantly stronger for intermediate attractive than for comparable repulsive scattering lengths. This suggests a fundamental difference in the underlying mechanisms on either side. The behavior in the vicinity of the attractive background scattering length as we observe it is similar to earlier measurements \cite{GUE06, OSP06B}.
Towards very strong attraction, we find a significant loss of $^{87}$Rb atoms, accompanied by an increase of visibility.
For strongly repulsive interactions, the visibility remains almost constant on a high level. This observation is consistent with phase separation in the lattice between bosons and fermions, which might be promoted by the slightly different trap shapes they experience, or the formation of a strongly anti-correlated mixed phase. Presently, we can not distinguish between both scenarios. For $a_{bf} > 400\,a_0$, we observe significant atom losses, however, these are essentially independent of the hold time, suggesting their occurance early during the lattice loading, when the mixture is held just below the FBR.

Based on the observations outlined above, we qualitatively divide the diagram  of Fig.~\ref{FIG1} into five distinct regimes, i.e. lossless coexistence in very shallow lattices for all scattering lengths, and regimes characterized by very strong interspecies attraction, intermediate attraction, weak interaction (of either sign), and very strong repulsion for deeper lattices. This classification also holds for low and intermediate values of $N_f$, although the symmetric feature around $a_{bf}=0$ is most pronounced for the highest $^{40}$K numbers.

We have verified that the visibility decrease is reversible for interspecies scattering lengths in the range $-200 \,a_0\lesssim a_{bf} \lesssim 400\,a_0$,
when $a_{bf}$ is tuned back to zero, provided that the system is given enough time to reequilibrate.
This suggests that the reduction of coherence is  caused by a reversible redistribution of the atomic densities during the lattice loading.
\begin{figure}
\includegraphics{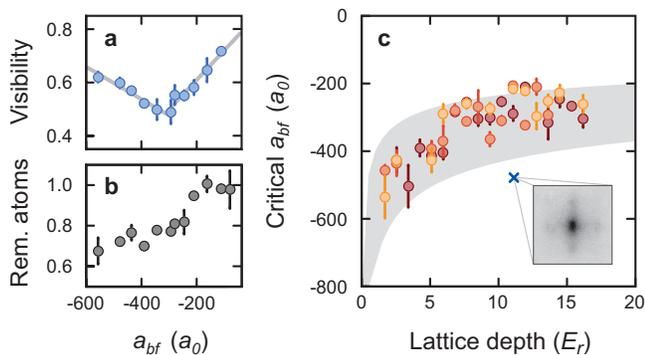}
\caption{\label{FIG3} Onset of the loss-dominated regime. (a) Experimentally determined visibility minimum and (b) simultaneous $^{87}$Rb loss feature at $V_0=9\,E_r$ for intermediate $N_f$. (c) Critical $a_{bf}$ determined from the visibility data, versus lattice depth for high, intermediate and low $N_f$ (darker colors for higher $N_f$). The shaded area gives an estimate of the scattering length at which the bosonic three-body loss timescale becomes shorter than the experimental hold time, based on our variational model. The inset shows a typical TOF image of $^{87}$Rb atoms in the loss-dominated regime, for low $N_f$. The cross indicates the associated experimental parameters ($V_0=11\,E_r$, $a_{bf}=-480\,a_0$). }
\end{figure}
For attractive scattering lengths beyond a critical value, we observe significant losses of $^{87}$Rb atoms, along with an increase in visibility. Despite this increase, no condensate is discernible in the corresponding TOF images. These findings are shown exemplarily in Fig.~\ref{FIG3}(a), (b). We fit two straight lines to the relevant part of the visibility profiles, and use their intersection point as an estimate for the critical value. Fig.~\ref{FIG3}(c) shows that the resulting critical scattering lengths are independent of $N_f$. This suggests that the visibility increase at strong attractions is governed by the onsite physics. Recently, the crucial role of the modification of the Wannier orbitals due to the mean-field onsite interaction in understanding the attractively interacting mixture has been pointed out \cite{LUE07}. A strong interspecies attraction will not only enforce high boson occupation numbers, but will also at given occupation yield increased onsite densities due to admixture of higher bands, which in turn could lead to a significant increase in three-body loss rates. 
We estimate this onsite density increase from a variational harmonic oscillator model, minimizing the total onsite energy with respect to the width $\sigma_b$ and $\sigma_f$ of the bosonic and fermionic Gaussian density profiles for given lattice depth and occupation numbers. For all practically relevant boson numbers, we find a critical scattering length at which the onsite density collapses. The strong density increase in the vicinity of this collapse gives rise to an enhanced bosonic three-body loss rate $\dot{N}_3$ due to the $\dot{N}_3 \propto \sigma_b^{-6}$ scaling.
For practical purposes, a site can be considered lost when $\dot{N}_3\,\tau \gg 1$, where $\tau$ is the experimental hold time. Assuming a loss coefficient for collisions of three $^{87}$Rb atoms in the range $K_3=4\ldots 7 \times 10^{-30}\,\mathrm{cm}^6\,/\mathrm{s}$  \cite{BUR97}, we find that lattice sites with four or more bosons will be lost on time scales faster than our hold time even in the absence of fermions. On the other hand, the observed ratio of lost $^{87}$Rb to $^{40}$K atoms between $3$ and $4$ suggests that three-body losses of two bosons together with one fermion do not play a significant role in our system. This observation is consistent with the measurements presented in \cite{GUE06}. 
Therefore, we focus our attention on sites occupied by three bosons and one fermion. 
The critical scattering length for bosonic three-body losses derived from our variational model shows good agreement with the experimentally observed onset of the loss-dominated regime, as can be seen in Fig.~\ref{FIG3}(c). We therefore conclude that for strong attraction, the dominant process is the loss of highly occupied sites enhanced by the interaction-induced density accumulation, commonly referred to as self-trapping \cite{LUE07, BRU08}. The associated increase of the visibility could be thought of as the removal of strongly localized atoms on highly occupied sites from the system, and thus a reduction of the bosonic filling in the system.

In order to investigate the persistence of bosonic superfluidity in the mixture, we extract the condensate fraction from the recorded TOF images. This is done by fitting anisotropic two-dimensional bimodal Gaussians to all first-order diffraction features as well as the central peak after substraction of a broad Gaussian background from the image.
The condensate fraction, which we define as the sum of the atom numbers in the narrow features at quasi-momentum zero and the eight first-order peaks, divided by the total atom number \footnote{A similar evaluation procedure has recently been used to study the bosonic superfluid to MI transition in two dimensions, see \cite{SPI08}.}, is found to decay monotonically towards zero with increasing lattice depth for all values of $a_{bf}$ at all fermion numbers.

\begin{figure}[t!]
\includegraphics[scale=1.0]{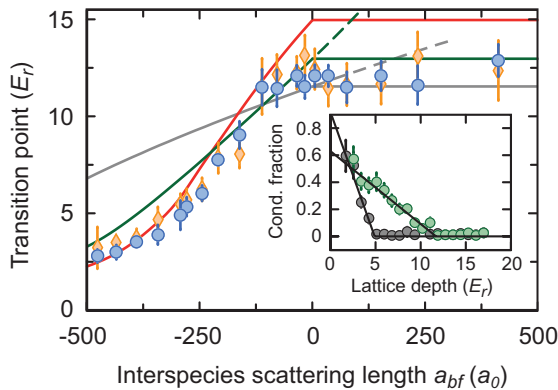}
\protect\caption{\label{FIG4} Shift of MI transition point with interspecies interaction. The diamonds and circles represent experimental runs at medium and high fermion numbers, respectively. The error bars incorporate the fit uncertainty and the weighted average over typically two experimental runs.  The lines indicate the theoretical MI transition point as estimated by the variational model for filling one (solid gray line), as well as exact diagonalization predictions for MI fillings of two (solid red) and one (solid green). For $a_{bf} > 0$, the solid lines correspond to a phase separation scenario, while the respective mixed phase predictions are depicted by dashed lines. The inset exemplarily shows the behavior of the condensate fraction at intermediate $N_f$, for $a_{bf}=-295\,a_0$ (black points) and $a_{bf}=235\,a_0$ (green points), respectively.}
\end{figure}

The point in lattice depth at which the condensate fraction vanishes, depends strongly on the interaction strength (Fig.~\ref{FIG4}). To quantify this point, we fit a linear decrease model to our data, thus obtaining the kink position for each scattering length (inset of Fig.~\ref{FIG4}). For medium to high $N_f$, the transition points lie almost on top of each other, both displaying a strong shift of up to $10\,E_r$ towards shallower lattices for attractive interactions. In this regime, we expect the bosons to move on top of an essentially homogeneous fermion filling. The fermions then constitute an effective attractive potential for the bosons, which adds to the depth of the lattice potential. The shift can thus be understood in terms of an effective deepening of the lattice sites induced by the Bose-Fermi attraction, which modifies both the boson tunneling and interbosonic repulsion. As soon as the interspecies attraction overcomes the bosonic repulsion, the formation of a bosonic MI shell of filling two is energetically favored, resulting in even deeper effective potentials, thereby further enhancing the modification of the Hubbard parameters. The variational model (gray line in Fig.~\ref{FIG4}) clearly underestimates the self-trapping and thus yields smaller shifts than experimentally observed. A single-site exact diagonalization approach for the effective Hubbard parameters including the full orbital degrees of freedom \cite{LUE07} (solid red and green lines in Fig.~\ref{FIG4}) shows good agreement with the experimental data \footnote{\criticalratio}. This supports that the shift of the transition point in the attractive regime can be explained by self-trapping. For repulsive interactions, a shift towards deeper lattices of comparable magnitude is predicted by the same model, assuming a homogeneous distribution of bosons and fermions throughout the lattice. The absence of a significant shift in this direction indicates that bosons and fermions do not occupy the same lattice sites in that regime. For low $N_f$, where the assumption of a homogeneous fermion filling is not valid, the decrease of the condensate fraction shows a more complex behavior which can not be captured by a linear slope, and the discussion of which is beyond the scope of our model.

In conclusion, we have explored the quantum phases of a mixture of bosonic $^{87}$Rb and fermionic $^{40}$K in a 3d optical lattice. By tuning the interspecies interaction and lattice depth, we have observed transparency of the fermions at $a_{bf}=0$, and a symmetric decay of visibility in a small window around that point. For stronger interactions, we find a remarkable asymmetry between the attractive and repulsive side. The condensate fraction shows a behavior which we interpret as a shift of the bosonic MI transition towards shallower lattice depths with increasing interspecies attraction. We attribute this shift to a renormalization of the parameters of the effective Bose-Hubbard model by self-trapping, and find good quantitative agreement with an exact diagonalization calculation.
Our results suggest that Feshbach-assisted lattice loading can be used to tailor the resulting lattice filling. Finally, the observed transparency is a promising starting point for experiments on tunable impurities \cite{GAV05, KRU08}.

\begin{acknowledgments}
We thank T.~Rom for his contributions to the experimental setup, and F.~Gerbier, A.~Mering, D.~Pfannkuche, M.~Snoek and M.~Zwierlein for valuable discussions. We acknowledge financial support by DFG, EU (IP SCALA, MC-EXT QUASICOMBS),
and DARPA (OLE). S.~W. acknowledges support by MATCOR.
\end{acknowledgments}


\end{document}